\documentclass[prb, twocolumn, letter, showpacs, superscriptaddress]{revtex4}
\usepackage{amssymb}
\usepackage{amsmath}
\usepackage{bbm}
\usepackage{graphicx}
\usepackage{color}

\begin{document}

\author{Ching-Kit Chan}
\affiliation{Columbia University, 538 West, 120th Street, New York, NY 10027, USA}
\author{Philipp Werner}
\affiliation{Theoretische Physik, ETH Zurich, 8093 Zurich, Switzerland}
\author{Andrew J. Millis}
\affiliation{Columbia University, 538 West, 120th Street, New York, NY 10027, USA}
\title{Magnetism and orbital-ordering in an interacting three band model: a dynamical mean field study }
\date{\today}

\hyphenation{}

\begin{abstract}
Single-site dynamical mean field theory is used to determine the magnetic and orbital-ordering phase diagram for  a model of electrons moving on a lattice with three orbital states per site and with the fully rotationally invariant Slater-Kanamori  on-site interactions. The model captures important aspects of the physics of transition metal oxides with partially filled $t_{2g}$ shells, and of  electron-doped C$_{60}$. We introduce an unbiased, computationally simple and inexpensive method for estimating the presence of two sublattice order, determine the regimes in which spatially uniform and two-sublattice spin and orbital orderings are present and give physical arguments for the origins of the different phases.   Guidelines are determined for optimizing the presence of ferromagnetism, which may be desirable in applications.
\end{abstract}


\maketitle

\section{Introduction}
Orbital degeneracy is believed to play a crucial role in the physics of many ``strongly correlated" materials\cite{Imada98} of current interest,  including ``early"  transition metal oxide compounds such as the lanthanum/strontium titanates \cite{Pavarini04} and vanadates, \cite{Mikokawa96,Sawada96,Fujioka08} ``late" transition metal oxide compounds such as the Sr/Ca ruthenate series,\cite{Liebsch03,Moore07,Liebsch07} and non-transition metal oxide compounds such as the  $A_n$C$_{60}$ series.\cite{Gunnarsson96} An important aspect of the physics of these compounds is the interplay between electron itineracy and the  rich multiplet structure arising from  the projection of the Coulomb interaction onto the orbitally degenerate on-site subspace.

The single-site dynamical mean field approximation \cite{Georges96} has had great success in treating the physics of materials in which only a single orbital is relevant. However, technical complications associated with the proper treatment of the full multiplet interactions have until recently  caused difficulties in the application of this formalism to the multiorbital case. The single-band applications were often based on the use of the Hirsch-Fye quantum Monte Carlo method\cite{Hirsch86} which does not have a straightforward generalization to the interactions present in the multiorbital case. A variant of the Hirsch-Fye method involving multiple auxiliary fields has been developed; however this method encounters a severe sign problem and difficulties with preserving rotational invariance in practice.\cite{Sakai04} Most published studies therefore involved approximations in which  the exchange and pair hopping terms of the full Coulomb interaction were either neglected, treated via uncontrolled analytical approximations \cite{Florens04,deMedici05,Lombardo05} or approximated in a way which breaks the rotational invariance.\cite{Held98}

The recent development of continuous-time quantum Monte Carlo methods \cite{Rubtsov05,Werner06a,Werner06b} along with improvements in the exact diagonalization technique \cite{Liebsch07,Liebsch07b,deMedici08} have made feasible a comprehensive theoretical treatment of multiorbital models  with the fully rotationally invariant interactions.\cite{Werner07,Werner08a,Werner09}

In this paper we use single site dynamical mean field techniques to determine the orbital and magnetic ordering phase diagram of a three orbital model. The work is an extension of a previous study of the paramagnetic phases of the model.\cite{Werner09}  We show that some but not all of the Mott transitions are preempted by magnetic and/or orbital ordering transitions, determine the relation of the magnetic phase diagram to the non-fermi-liquid instability discussed in previous work\cite{Werner08a} and demonstrate that for the model we study ferromagnetism exists only at carrier concentrations $n>1$ and for very strong correlations. On the technical side we introduce a computationally inexpensive method to estimate the  location of the phase boundary separating phases with uniform or two sublattice orbital and magnetic order from phases with no long ranged order, and where the order occurs estimate the transition temperatures.

The rest of this paper is organized as follows. Section \ref{methods} presents the model to be studied and the methods we use, including the simple approach for identifying phase boundaries, Section \ref{phase} presents the phase diagram and gives physical arguments elucidating the origin of the various ordered phases, Section \ref{Tdependence} presents some results on the temperature dependence of phase boundaries and Section \ref{Conclusion} is a summary and conclusion, outlining the implications of our results for experiments and prospects for future work.


\section{{Model and methods}\label{methods}}
\subsection{Model}
We analyze the ``three band" model defined by the Hamiltonian
\begin{equation}
H=H_\text{band}+H_\text{int}
\label{H}
\end{equation}
with
\begin{eqnarray}
H_\text{band}&=&-\sum_{\langle ij \rangle \alpha\beta\sigma}t^{\alpha\beta}_{ij}\psi^{\dagger}_{i\alpha\sigma}\psi_{j\beta\sigma}-\sum_{i\alpha\sigma}\mu n_{i\alpha\sigma}
\label{Hband}\\
H_\text{int}&=&\frac{1}{2}\sum_{i\alpha\beta\sigma\sigma'}U_{\alpha\beta\sigma\sigma'} n_{i\alpha\sigma} n_{i\beta\sigma'}\nonumber
\label{Hint}
\\
&-&J\sum_{i,\alpha\ne\beta}(\psi^\dagger_{i\alpha\downarrow}\psi^\dagger_{i\beta\uparrow}\psi_{i\beta\downarrow}\psi_{i\alpha\uparrow}
+ \psi^\dagger_{i\beta\uparrow}\psi^\dagger_{i\beta\downarrow}\psi_{i\alpha\uparrow}\psi_{i\alpha\downarrow}  \nonumber \\
&& \ \ \ \ \ \ \ \ \ \  + h.c.).
\end{eqnarray}
$H_\text{band}$ includes the usual electronic hopping between orbital $\alpha=1,2,3$ on site $i$ and orbital $\beta$ on site $j$, and the chemical potential $\mu$.  The fourier transform of $t^{\alpha\beta}_{ij}$ is a dispersion ${\bf E}(p)$ which is a matrix in orbital space. The two terms in Eq.~(\ref{Hint}) are the on-site interaction. The form of this term  follows from symmetry considerations. The various $U_{\alpha\beta\sigma\sigma'}$ terms are equal to (i) $U$ when $\alpha=\beta,\sigma\neq\sigma'$, (ii) $U'=U-2J$ when $\alpha\neq\beta,\sigma\neq\sigma'$ and (iii) $U'-J=U-3J$ when $\alpha\neq\beta,\sigma=\sigma'$; where $U$($U'$) accounts for intra(inter)-orbital Coulomb repulsion among orbitals $\alpha$ and $\beta$. $J$ in the last term is the coefficient of the spin exchange and pair hopping terms. We shall be interested in cases in which the point group symmetry of the lattice guarantees that (in the absence of spontaneous symmetry breaking) the on-site Green function $G^{\alpha\beta}(R=0,\omega)\sim \delta_{\alpha\beta}$; this condition is satisfied in the materials listed in the Introduction.

\subsection{Dynamical Mean Field Approximation}

To solve Eq.~(\ref{H}) we use the single-site dynamical mean field method \cite{Georges96} which neglects the momentum dependence of the self-energy. To study magnetic or orbital orderings which spontaneously break the translational symmetry of the lattice down to a lower symmetry characterized by several sublattices,  one must in principle  introduce a quantum impurity model for each sublattice and take the self energy for the lattice problem to be local, but different on each sublattice. For simplicity we focus here on two sublattice orderings. We may then associate a sublattice index $\lambda=e$ or $o$  to the  self energy (which also may be a matrix in spin and orbital space), distinguish hopping between the same and different sublattices and write the lattice Green function as a two by two matrix in sublattice space
\begin{equation}
{\bf G}^{-1}=\left(\begin{array}{cc}\omega-{\bf E}_\text{same}(p)-{\bf \Sigma}_e(\omega) &{\bf  E}_\text{diff}(p) \\{\bf E}_\text{diff}(p) & \omega-{\bf E}_\text{same}(p)-{\bf \Sigma}_o(\omega)\end{array}\right).
\label{G2sublattice}
\end{equation}
We assume that in the absence of spontaneous symmetry breaking (i.e. if ${\bf \Sigma}_e={\bf \Sigma}_o\sim \delta^{\alpha\beta}_{\sigma_1\sigma_2}$) the dispersions ${\bf E}_\text{same,diff}$ are such that the local Green function
\begin{equation}{\bf G}_\text{loc}(\omega)=\int(dp) {\bf G}(p,\omega)
\label{Gloc}
\end{equation}
is proportional to the unit matrix in orbital space.

The self energies ${\bf \Sigma}_{e,o}$ are obtained from the solution of  quantum impurity models (one for each sublattice $\lambda$) of the form
\begin{equation}
H_{QI;\lambda} =-\sum_{\alpha, \sigma}(\mu-\Delta_{\alpha\sigma\lambda}) n_{\alpha,\sigma,\lambda} +H_\text{loc}+H_\text{hyb}+H_\text{bath},
\label{HQI}
\end{equation}
with $n_{\alpha,\sigma,\lambda}$ the density of electrons of spin $\sigma$ on orbital $\alpha$ in the model pertaining to sublattice $\lambda$ and $\mu$ and $\Delta$ encoding the chemical potential and any ligand fields arising from the explicit breaking of the point group symmetry.  Spontaneous symmetry breaking is signaled by the appearance of a spin, orbital or sublattice-dependent Hartree term in the self energy. The interaction term $H_\text{loc}$ consists of the on-site interaction terms of the original model. The remaining terms, $H_\text{hyb}$ and $H_\text{bath}$ are bilinear in fermion operators and produce a hybridization function ${\bf F}^\lambda(\omega)$ which is a matrix in spin and orbital space and whose form is  fixed by the self consistency condition that the Green function of the quantum impurity model pertaining to sublattice $\lambda$, ${\bf G}^{QI;\lambda}$,  be equal to the $\lambda-\lambda$ component of ${\bf G}_\text{loc}$ (Eq.~(\ref{Gloc})): ${\mathbf G}^{QI;\lambda}={\mathbf G}_\text{loc}^{\lambda\lambda}$.

In this paper we employ the semicircular density of states  $\rho(\epsilon)=\sqrt{4t^2-\epsilon^2}/(2\pi t^2)$, corresponding to a Bethe lattice with infinite coordination number  and a fully bipartite hopping Hamiltonian (${\bf E}_\text{same}$ in Eq.~(\ref{G2sublattice}) $\equiv0$). For this case the self-consistency equation in the general two-sublattice case becomes
\begin{eqnarray}
{\cal F}_{A,\alpha,\sigma}(-\omega)=-t^2G_{B,\alpha,\sigma}(\omega)
\label{bethesce}
\end{eqnarray}
with ${\cal F}_{A,\alpha,\sigma}$  the hybridization function
for sublattice A with orbital $\alpha$ and spin $\sigma$ and $G_{B,\alpha,\sigma}$ the local green function on the other sublattice.\cite{Werner06b} The self consistency equation for the  translation-symmetry unbroken case is obtained by setting $G_B\rightarrow G_A$ in Eq.~(\ref{bethesce}) and not considering the B sublattice.
We solve the quantum impurity model using the methods introduced in Refs.~\onlinecite{Werner06a,Werner06b}. 
The self-consistency condition is solved by iteration.

For comparison we have also solved the model in the Hartree Fock approximation using standard methods. For this problem the Hartree-Fock approximation is equivalent to replacing the full dynamical mean field self energy by
\begin{eqnarray}
\Sigma_{HF}^{\alpha\sigma}&=&U\langle n_{\alpha{\bar \sigma}}\rangle\nonumber \\
&+&\sum_{\beta\neq \alpha}\Big\{(U-2J)\langle n_{\beta{\bar
\sigma}}\rangle +(U-3J)\langle n_{\beta{\sigma}}\rangle\Big\} \label{SigmaHF} \nonumber \\
\end{eqnarray}
with ${\bar \sigma}$ the spin direction opposite to $\sigma$ and $\alpha,\beta$ labeling orbitals. The expectation values are
determined self-consistently. The solution of the Hartree-Fock equations is simplified by the observation that the system becomes fully spin polarized before orbital ordering occurs.

\subsection{Method of determining two sublattice ordering}

\begin{figure}[t]
\begin{center}
\begin{tabular}{cc}
\includegraphics[angle=-90, width=1.0\columnwidth]{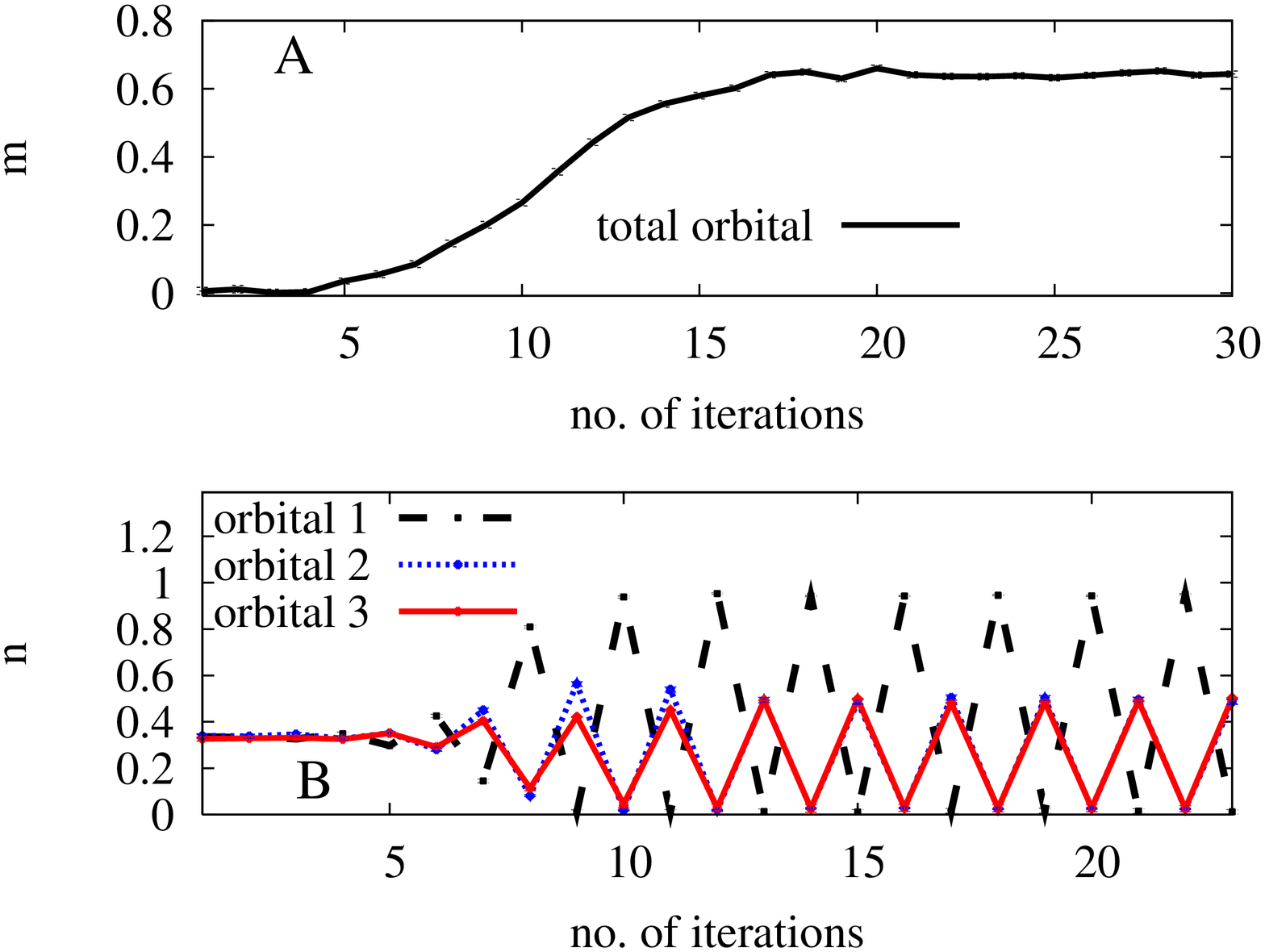}\\
\includegraphics[angle=-90, width=1.0\columnwidth]{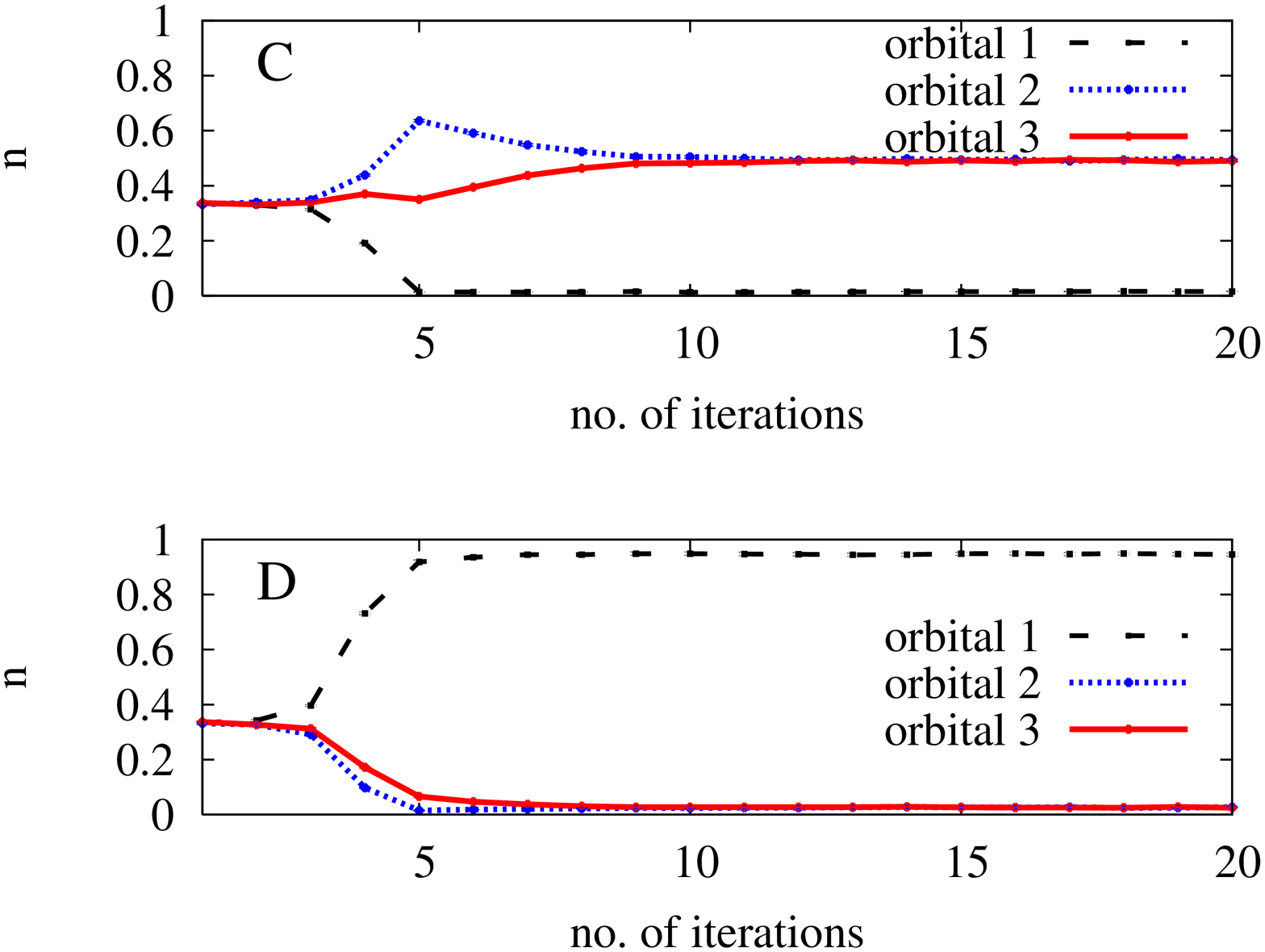}
\end{tabular}
\caption{Variation of the on-site orbital occupancy ($n$) and magnetization ($m$) with number of iterations of the dynamical mean field self consistency condition, computed for a semicircular density of states at density $n=1$ with $\beta t=50$, $U/t=16$ and  $J=U/6$.  Panel (A): iteration number dependence of the magnetization (summed over all orbitals); saturation indicates a ferromagnetic state with moment $m\approx 0.65n$. Panel (B): iteration number dependence of occupancies of the three orbitals, obtained from a translation-symmetry-unbroken (Eq.~(\ref{bethesce}) with $G_B\rightarrow G_A$) dynamical mean field equation; oscillations of the occupancy indicate two sublattice orbital ordering.  Panels (C) and (D): iteration number dependence of occupancies of the three orbitals for sublattice A (panel C) and sublattice B (panel D), obtained from the general self-consistent condition Eq.~(\ref{bethesce}); saturation of occupancies to a two-sublattice orbital state is evident. Error bars have the same size as the data points. }
\label{orbital}
\end{center}
\end{figure}

Finding orbitally and magnetically ordered solutions involves solving several quantum impurity models; it is therefore computationally more expensive than studying the paramagnetic case; in addition one must (as in any mean field theory) choose which symmetry breaking to investigate.  We have found, however, that  the presence of two sublattice order is indicated, to a good approximation and in an unbiased way, by an oscillatory behavior in the solution of the  dynamical mean field equations using the self-consistency condition appropriate for the case with unbroken translational symmetry.  In studies of the symmetry-unbroken phases of quantum impurity models it is common practice to enforce the lack of order and improve the statistical accuracy by symmetrizing the hybridization function at each iteration of the self-consistency equation. However, if this is not done, then the dynamical mean field self consistency procedure will sometimes fail to converge, exhibiting instead an oscillating behavior which, we argue,  signals the presence of two sublattice ordering.

Figure~\ref{orbital} presents a particular example: results of a dynamical mean field solution of the three orbital model with semicircular density of states, interaction strength   $U/t=16$, inverse temperature  $\beta t=50$, and density $n=1$. For these parameters the model is in its Mott insulating regime, and as will be seen has ferromagnetic spin ordering and two-sublattice orbital ordering. The dynamical mean field equations are solved by iteration from a non-ordered seed state and the results are plotted against iteration number.  Figure~\ref{orbital}A  presents the magnetization summed over all orbitals (the result is essentially the same whether the full Eq.~(\ref{bethesce}) or its translation-symmetry-unbroken special case is used). One sees the magnetization increase and saturate.  Figure~\ref{orbital}B presents results obtained for the occupancies of the individual orbitals using the version of Eq.~(\ref{bethesce}) in which the translational symmetry breaking is not allowed ($G_B\rightarrow G_A$). One sees that a stable solution is not obtained. Instead, there is an oscillation  between two states with different orbital occupancies.   To demonstrate the meaning of this oscillation we present in panels C and D results obtained from a solution of Eq.~(\ref{bethesce}), which explicitly allows for two sublattice translational symmetry breaking.   Panel C shows orbital occupancies corresponding to one sublattice and panel D to the other sublattice. Comparison to panel B shows that the two states between which the self-consistent equations oscillate correspond to the states of the two sublattices in the ordered solution.  We have also investigated other parameters with similar results, although care is sometimes required because when the two sublattice equations are used with a symmetric initial condition, two sublattice order may sometimes require a large number of iterations to develop, either because of a small Lyapunov exponent describing the growth of the broken symmetry phase or because the symmetry unbroken state is locally stable, so that one must wait for a fluctuation which is large enough to activate the system over a barrier.

Of course, the existence of a  two-sublattice solution to the mean field equation which persists for many iterations indicates at best the presence of a locally stable phase, and the parameter at which such a solution appears or disappears may indicate a spinodal point,  as in many circumstances the transitions between phases are first order. An energy analysis would be required to determine the precise location of the phase boundaries, and the energy differences involved are often very small (for example, the energy of the $O1$ paramagnetic insulating state at $n=1$ and $U=16t$ is found to  differ from that of the paraorbital paramagnetic state by an amount of order  $0.3\%$ of the total energy).

The phase boundaries presented in the rest of this paper are obtained by determining the onset of oscillations in the solution of the translation-symmetry-unbroken equations and should be understood with these caveats.

\section{{Phase Diagram}\label{phase}}

\begin{figure}[b]
\begin{center}
\includegraphics[angle=-90, width=1\columnwidth]{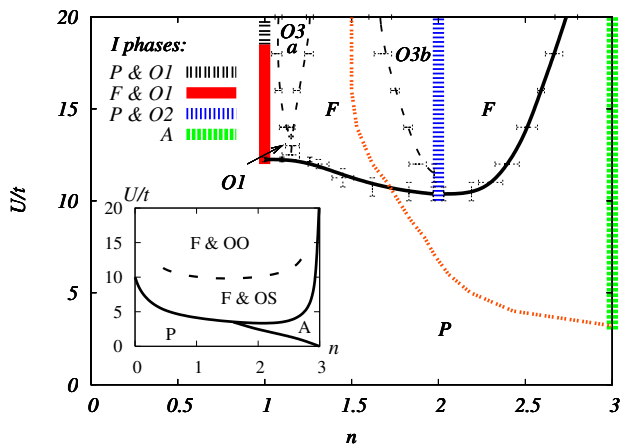}
\caption{(Color online) Main panel: phase diagram of the 3 band model with semicircular density of states at $\beta t=50$ and $J=U/6$ in the plane of particle density $n$ and interaction strength $U$. The vertical lines  indicate the Mott insulating
phases at integral values of $n$. The magnetic state is labeled by P (paramagnetic), F (ferromagnetic) and A (two sublattice
antiferromagnetic) while the labels O(N) denote  the 3 classes of orbital ordering discussed in the text. The heavy dashed line (orange on-line) gives the boundary of the non-fermi-liquid frozen-moment phase discovered in Ref.~\onlinecite{Werner08a}. Inset:  Hartree-Fock phase diagram for magnetic phases of  the same model. Magnetic phase boundaries are indicated by solid lines and orbital ordering boundaries by dashed lines. OO and OS stand for the orbitally-ordered and orbitally-symmetric phases respectively. All transitions are second order except the FM-AFM transition and
the orbital ordering transitions at $U\gtrsim 12t$ and small $n$.}
\label{phasediagram}
\end{center}
\end{figure}

\subsection{Overview}

Figure~\ref{phasediagram} presents the phase diagram in the plane of Coulomb repulsion and band filling  obtained using the methods described above with $J$ fixed to be $U/6$ and at the low temperature $\beta t=50$ (temperature $1/200$ of the bandwidth). The phase diagram exhibits insulating phases at integer band fillings $n=1,2,3$ and sufficiently strong interactions.  A new feature of the present paper relative to earlier work is the explicit inclusion of orbital and magnetic ordering. Comparison to Fig. 2 of Ref.~\onlinecite{Werner09} shows that the inclusion of ordering does not shift  the position of the Mott lobes appreciably: the critical $U$ at the tip of the $n=1,2$ lobes is essentially unchanged.

Away from integer doping the phases are metallic.  The various magnetically and orbitally ordered phases and their phase boundaries are indicated, as is the boundary to a phase discovered in earlier work \cite{Werner08a} that is  characterized by frozen moments and a non-fermi-liquid self energy.   We have made selective studies of other values of $J$ finding little impact on the qualitative features of the phase diagram. For example, for $n$ between $1$ and $2$, increasing $J$ to  $U/4$ shifts the magnetic phase boundary downwards in $U$ by about $0.5t$. The inset presents the Hartree-Fock phase diagram, which is discussed below.

\begin{figure}[t]
\begin{center}
\includegraphics[angle=-90, width=0.9\columnwidth]{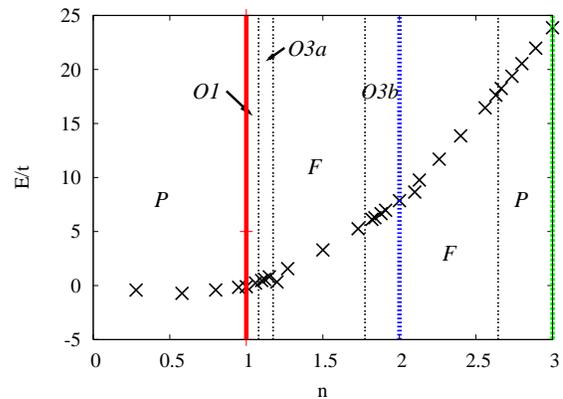}
\caption{Internal energy $E$ computed as a function of density per site $n$ at  $\beta t=50$, $U/t=16$. Vertical lines give the positions of phase boundaries presented in Fig.~\ref{phasediagram}}.
\label{ENERGY}
\end{center}
\end{figure}

The phases we find are apparently stable against phase separation. Figure~\ref{ENERGY} shows the energy computed as a function of density for $U=16t$; the upward concavity required for stability against phase separation  is evident.

\subsection{Magnetic phases}

In the density range $0<n<1$ we found no evidence for magnetic order in the range of interaction strength studied. Ferromagnetic phases do occur at $n=1$ (although for the largest $U$ studied the Curie temperature falls below $\beta t=50$ so no order is detected in our calculations). Ferromagnetic phases also occur for  $1<n<2$. At $n=2$ any uniform or two-sublattice  magnetic transition is below the temperature $\beta t=50$ of this study.  We suspect that at $n=2$ the magnetic  order is actually incommensurate or has a higher order commensurability as was found for intermediate couplings in other models with non-integer number of electrons per orbital per site.\cite{Chattopadhyay01} Ferromagnetism is also found for a range of $n$ between $2$ and $3$ while a two-sublattice antiferromagnetic state exists for $n=3$.

Note that the perfect nesting of the model we study implies that at half filling ($n=3$) the antiferromagnetic phase is the ground state for all $U>0$. The small $U$ limit of the antiferromagnetic phase marks the point at which the Neel temperature falls below the temperature $T=t/50$ used in the computations.

A frozen-moment phase was recently reported by two of us in the three-orbital model.\cite{Werner08a} The boundary to this phase is shown as the dotted line (brown on-line) in Fig.~\ref{phasediagram}. Our results indicate that the frozen-moment phase is preempted by the ferromagnetic phase for strong correlations but that the frozen moment phase (which presumably indicates a non-uniform magnetic state; either disordered or a long-period spiral) exists over a wide range of dopings at weaker correlations.

Remarkably, in the DMFT calculations the ferromagnetic phase appears  to be confined to interaction strengths of the order of or larger than the critical value for Mott insulating behavior and to densities greater than or equal to $1$.   Changing the Hunds coupling $J$ over the physically relevant range $J<U/3$ does not change the results significantly.  The DMFT results are in sharp contrast to the Hartree-Fock results shown in the inset of Fig.~\ref{phasediagram} which indicate a much wider range of ferromagnetism and orbital order, extending in particular to much lower $U$ and to densities $n<1$. Our results suggest that LDA+$U$ band theory calculations, which are in essence a Hartree-Fock approximation to the strong on-site interactions, may severely overestimate the range in which ferromagnetism and orbital order occurs.

The absence of ferromagnetism at carrier concentrations $n<1$ is similar to  results obtained for two-orbital models by Momoi and Kubo \cite{Momoi98} using a dynamical mean field method with an exact diagonalization solver and Held and Vollhardt \cite{Held98} using Hirsch-Fye QMC  with only the Ising component of the Hunds interaction retained, and also to very recent Gutzwiller approximation calculations \cite{Buenemann09} which however yield a much wider range of ferromagnetism at $n>1$ than is found in our calculations.

\begin{center}
\begin{table}[htb]
\begin{tabular}{|p{1.25cm}|c|c|}\hline
Types of ordering&  $(n^A_{\alpha},n^A_{\beta},n^A_{\gamma}|
n^B_{\alpha},n^B_{\beta},n^B_{\gamma})$  & Symbols
\\ \hline
O1&  \scriptsize$(1-\delta,\delta,\delta|\delta,{1\over2},{1\over2})$   &
(\LARGE$\uparrow$\:\tiny$\uparrow$\;\;\:\tiny$\uparrow$\normalsize\;\huge$|$\large\:\tiny$\uparrow$\;\;\normalsize$\uparrow$\;\normalsize$\uparrow$\small)
\\ \hline
O2& \scriptsize$({1\over2},{1\over2}, 1|1,1,0)$    &
(\normalsize$\uparrow$\:\;\normalsize$\uparrow$\;\huge$\uparrow$\huge$|$\huge$\uparrow$\huge$\uparrow$\small\;0)
\\ \hline
O3a& \scriptsize$(1-2\delta,2\delta,\delta|\delta,2\delta,1-2\delta)$    &
(\LARGE$\uparrow$\;\scriptsize$\uparrow$\:\,\tiny$\uparrow$\Large\;\huge$|$\large\;\tiny$\uparrow$\;\;\;\scriptsize$\uparrow$\;\LARGE$\uparrow$\small)
\\ \hline
O3b& \scriptsize$(2\delta,1-2\delta,1-\delta|1-\delta,1-2\delta,2\delta)$    &
(\scriptsize$\uparrow$\tiny\:\:\LARGE$\uparrow$\huge$\uparrow$\tiny\;\huge$|$\huge$\uparrow$\LARGE$\uparrow$\small\;\scriptsize$\uparrow$\;\small)
\\ \hline

\end{tabular}
\caption{Characterization of orbital orders.$n_{\alpha}^A$ gives the density of electrons of the $\alpha$ orbit in sublattice A, etc.$\delta$ is the deviation of the total density n from the integer value   1  for O1 and O3a or 2 (O3b). Arrow lengths indicate magnitudes of densities.}
\label{table1}
\end{table}
\end{center}

\subsection{Orbital Ordering}
We now turn to the complex orbital ordering phase diagram revealed in Fig.~\ref{phasediagram}. All the $d$-orbital orderings predicted by our DMFT calculations are staggered, rather than homogeneous in space. The nature of the phases is explicated in Table \ref{table1}, but the information about occupancies must be interpreted with care. For example, minimizing the interaction energy at density $n=1$ leads to a two sublattice ordered phase. In this phase, one sublattice has  one orbital (say orbital $1$) occupied and the other two ($2$ and $3$) empty, while in the alternate sublattice orbital $1$  is empty and orbitals $2$ and $3$ are half filled. We believe the correct physical interpretation of the half filled state is that it represents an incoherent superposition of the state with orbital $2$ filled and $3$ empty and the reverse, so that in the lattice the state corresponds to a highly degenerate set of states in which half of the sites in one sublattice have  orbital $2$ occupied while the other half of the sites on this sublattice have orbital $3$ occupied.  Supporting evidence for this interpretation comes from the measurement of the $\langle n_2 n_3 \rangle$ equal-time correlation function. At carrier concentration $n=1$ it is found to be very small (of order $10^{-4}$).

This behavior can also be understood by considering the strong coupling limit.   If we assume a fully spin polarized ferromagnetic state then the strong coupling Hamiltonian describing the orbital ordering is (to leading order in $t/U$)  an antiferromagnetic three state Potts model with nearest-neighbor interactions. On the cubic lattice this model is known to have a low temperature phase with precisely this structure.\cite{Banavar82}  It is remarkable that this nontrivial state is correctly identified by our simple ``find the oscillations in the DMFT iteration" procedure. We observe that considering higher orders in $t/U$ would lead to longer ranged interactions which would lift the degeneracy, leading to ordered states characterized by larger unit cells beyond the scope of our calculation. We suggest that the four-sublattice state discussed for LaTiO$_3$ by Pavarini {\it et al.}\cite{Pavarini04} is of this type.

Doping changes the nature of the phase, converting the second sublattice from gapped to metallic (albeit with a low fermi-liquid temperature) in which the low $T$ phase is (on long timescales) a coherent combination with each of the two sites half-occupied. In this phase the equal time $\langle n_2 n_3 \rangle$ correlation function is larger, of order $10^{-2}$. This phase, which is the stable solution of the DMFT equations, has within our $0.1\%$ accuracy  the same energy as a fully
orbitally symmetric weakly ferromagnetic metal phase and it is possible that this latter phase (which we typically find to be
unstable to the O1 phase) is the true ground state for some dopings.

As seen in Fig.~\ref{phasediagram}, further doping produces a transition to the $O3a$ state. We have verified that raising the temperature shifts the phase boundary between $O1$ and $O3a$ near $n=1$  to the right as expected if $O1$ is the high entropy Potts model state whereas $O3a$ is a lower entropy state. Similar arguments also apply for the competing orders $O2$ and $O3b$. However, the situation near $n=2$ differs in two important respects from that near $n=1$: the $O2$ phase seems not to survive even an infinitesimal doping, and the $O2$ phase is entirely paramagnetic.

The trends in the phase boundaries suggest that at strong coupling the intermediate disordered phase separating the $O3a$ and $O3b$ phases will be eliminated. We observe that at $\delta=1/2$ the $O3a$ and $O3b$ phases become identical, implying that for large enough $U$ where the disordered phase vanishes,  the $O3a$ and $O3b$ states are adiabatically connected.

Finally, we observe that  the  Hartree-Fock approximation (inset to Fig. \ref{phasediagram}) predicts a transition to an orbitally ordered state. The Hartree-Fock transition to the orbitally ordered phase occurs at relatively large $U$, such that (within Hartree Fock) the spins are fully polarized. The transition is found to be second order except at very large $U\gtrsim 12t$ and small $n$, where it becomes first order. Again we regard the qualitative difference between the Hartree Fock and DMFT results as indicating that caution is necessary in interpreting LDA+$U$ calculations.

\section{{Temperature Dependence }\label{Tdependence}}
\begin{figure}[t]
\begin{center}
\includegraphics[angle=-90, width=1.0\columnwidth]{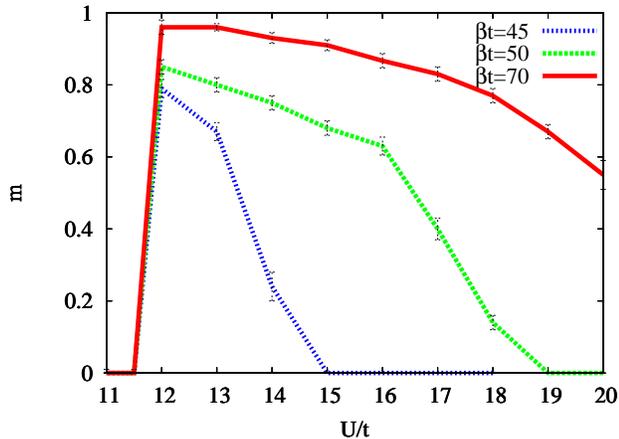}
\caption{Magnetization, $m$, plotted against interaction strength $U$ for different inverse temperatures $\beta$ at density $n=1$ indicating a  very rapid onset of magnetism as the material enters the Mott insulating phase and a weakly interaction-strength dependent Curie temperature at large $U$.
}
\label{N1}
\end{center}
\end{figure}

In this section we consider the temperature dependence of the order parameters and the nature of the thermally-driven phase transitions. Figure~\ref{phasediagram}  indicates that in the Mott insulating phase at $n=1$ and at our chosen temperature $\beta t=50$, the system exhibits a phase transition from the ferromagnetic to the paramagnetic phase as $U$ is increased above  $U\sim18 t$. To clarify the nature of this phase transition, we present in Fig. \ref{N1} the dependence of the magnetization on interaction strength  for different temperatures. The rapid and apparently $T$-independent onset of the magnetization at $U\sim 11.5t$ suggests  a first order magnetization onset which seems to coincide with the onset of the Mott insulating phase. More detailed studies are required to clarify the behavior near $U=11t$ precisely.  For the larger $U$-behavior, comparison of the curves corresponding to different temperatures reveals a weakly $U$ dependent Curie temperature, of the order of $0.022t$ at $U\sim 14.5t$, $0.02$ at $U=18.5t$ and decreasing slowly for larger $U$, consistent with the ferromagnetic superexchange $\sim t^2/U$ expected in this orbitally degenerate situation.

\begin{figure}[t]
\begin{center}
\includegraphics[angle=-90, width=1.0\columnwidth]{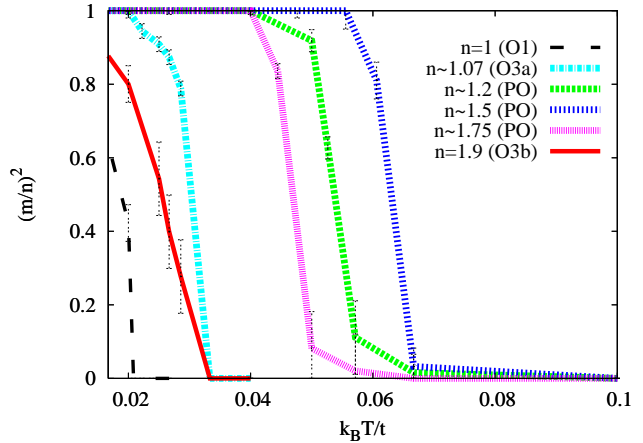}
\caption{Square of the magnetization (normalized to density) for various magnetic phases at $\beta t=50$ and $U/t=16$. the curve shown for density $n=1$ corresponds to an insulating phase, while all other phases are metallic. The paraorbital
phases (PO), $n\sim 1.2, 1.5, 1.75$ have higher Curie temperature ($k_B T/t \sim 0.05-0.07$) than the orbitally ordering phases. }
\label{m_T}
\end{center}
\end{figure}

To investigate the effect of doping we plot in Fig.~\ref{m_T}  the temperature dependence of the square of the magnetization, normalized to the particle density for various dopings between $1$ and $2$ at a $U$ somewhat larger than the Mott critical values. At low $T$ the spin polarization is complete ($m=n$) except perhaps in the $n=1$ insulating phase. One sees that metallic phases have higher Curie temperatures than insulating phases, and paraorbital phases higher Curie temperatures than orbitally ordered phases. The maximal value of the transition temperature is seen to occur about  half-way between the Mott lobes, i.e. at  $n\approx 1.5$, where  the model is maximally metallic. This finding is consistent with the argument of Momoi and Kubo \cite{Momoi98} that the physics of ferromagnetism in multiorbital models is related to the physics of `double exchange': the strong interaction and non-vanishing Hunds coupling puts each site into its maximal spin state and the ferromagnetic transition temperature is then determined by the energetics of carriers hopping in the locally spin-polarized background.

Figure~\ref{m_T} also indicates that except very near to the Mott insulating phases, the temperature-driven ferromagnetic transition is very steep, indeed within our accuracy  apparently discontinuous. It is interesting that apparently first order behavior is most pronounced for  dopings where there is no orbital order, while if orbital order is present the transitions seem more continuous.

\begin{figure}[t]
\begin{center}
\includegraphics[angle=-90, width=1.0\columnwidth]{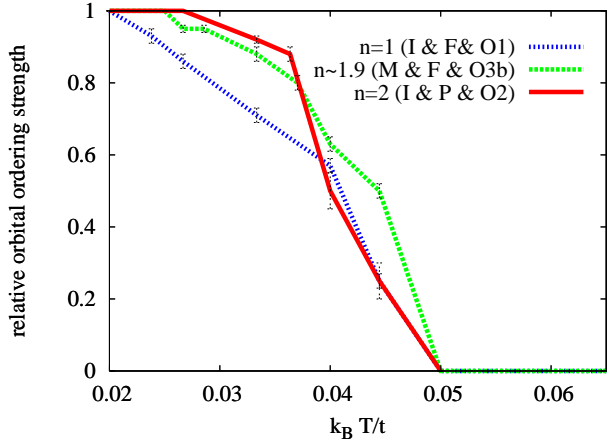}
\caption{Relative orbital ordering strength for various phases at $\beta t=50$ and $U/t=16$. ``Relative orbital ordering strength" is the deviation of the occupancy of one orbital from the paraorbital value ($n/3$) normalized to the value at our lowest measurement temperature $k_B T=0.02t$; mathematically  $\big(n^A_{\alpha}(k_B T>0.02t)-\frac{n}{3}\big)/\big(n^A_{\alpha}(k_B T=0.02t)-\frac{n}{3}\big)$. The transitions from the orbital selective to the paraorbital phase are apparently second order. The transition temperatures are not sensitive to the metallic (M/I), ferromagnetic (P/F) or ordering (O1/O2/O3) nature of the low $T$ phase.
}
\label{OO_T}
\end{center}
\end{figure}

Figure~\ref{OO_T} shows the temperature dependence of the order parameter for orbital ordering for several different carrier concentrations. The relative orbital ordering strength shown in the figure compares the deviation of $n_{\alpha}^A$ from $n/3$ (the paraorbital value) to the corresponding value at low temperature ($k_B T=0.02t$). The staggered orbital ordering has its critical temperature at around $k_B T/t\sim 0.04-0.05$ and neither the value of the transition temperature nor the order of the transition depends on whether the system is metallic or insulating, paramagnetic or magnetic. In the region where both orbital and magnetic ordering occurs the orbital ordering transition temperature is higher than the magnetic one.

\section{{Conclusions}\label{Conclusion}}
We have used the single-site dynamical mean field approximation to determine the magnetic and orbital-ordering phase diagram of a three-band model with realistic rotationally invariant multiplet interactions and have introduced a computationally simple and unbiased method for identifying the presence of two-sublattice long ranged order from an oscillation with iteration number in the symmetry-unbroken DMFT equations.  We find that the Mott insulating phases at band fillings $n=1$ and $n=2$ per site are unstable towards a two-sublattice orbital ordering of the degenerate type associated with the three-state Potts model on a bipartite lattice,\cite{Banavar82} while as expected at $n=3$ the Mott phase is pre-empted by Neel ordering. It is interesting that the nontrivial ground state is the one identified by the unbiased method of examining oscillations in the DMFT iterations.  Ferromagnetism occurs at $n=1$ and for $n$ between 1 and 2 and for carrier concentrations greater than $n=2$ but not too close to $n=3$. The onset temperatures for orbital and magnetic ordering are low,  of the order of $1$-$2\%$ of the full band width.

Remarkably, at $n=1$ and $n=2$ the orbital ordering is essentially co-terminus with the Mott phase: orbital ordering seems not to exist for interactions less than the Mott critical value, and only a tiny region of the Mott phases ($U$ within a few percent of $U_c$)  appears not to have orbital ordering.   A high precision study of the behavior for $U$ very close to $U_{c2}$, beyond the scope of this work,  would be required to verify the detailed behavior for $U\approx U_{c2}$. It is of course also possible that a longer period ordering (not considered in our work) would extend also into the metallic phases away from the Mott region. These issues, as well as the possibility of phase separation,  are important topics for further investigations.

Ferromagnetism is found only for large interaction strengths, greater than or of the order of the critical values needed to drive a Mott transition at $n=1,2$ and only for carrier concentrations greater than $n=1$ per site but not too close to $n=3$ per site. As also discussed by Momoi and Kubo, the physics of ferromagnetism in the multiorbital models appears to be related to the physics of `double-exchange' and orbital selectivity: at $n>1$ and strong correlations one orbital becomes occupied by one (spin-polarized) electron; this occupied orbital acts as a `core spin' whose orientation controls the hopping of the remaining electrons, leading to ferromagnetism similar to that in the manganites. For $n<1$ the core spin effect is absent, and we expect that the physics of ferromagnetism is similar to that discussed in the context of the one-orbital Hubbard model,\cite{Vollhardt97} where densities of states peaked in the band center (as is the case for the semicircle) disfavor ferromagnetism and densities of states peaked at the band edges favor it. We have found that varying the value of $J$ within a physically reasonable range does not change the phase boundaries appreciably. However, the considerations from the one-orbital Hubbard model suggest that even for $n>1$ variations in the density of states, in particular shifting the maximum away from the center of the band towards a band edge  may widen the range over which ferromagnetism exists.

The development of `oxide heterostructures' \cite{Mannhart08} has opened new avenues for material design.  In the context of the LaTiO$_3$/SrTiO$_3$ and LaAlO$_3$/SrTiO$_3$ heterostructures theoretical \cite{Okamoto06,Pentcheva07} and experimental \cite{Brinkmann07} reports of ferromagnetism have appeared. The materials of interest in these latter studies, however, were titanate-based, corresponding to densities $n<1$. The theoretical predictions of ferromagnetism and orbital ordering were based on the LDA+$U$ approximation, which is in effect a Hartree-Fock approximation to the local correlations. Our dynamical mean field results  cast doubt on these  theoretical predictions, and suggest  that the magnetic behavior observed in Ref.~\onlinecite{Brinkmann07} may be due to something other than an interface ferromagnetic phase.  Our results suggest that attempts to obtain a ferromagnetic electron gas at an oxide interface should focus on electron-doping a material with $n=1$ or hole-doping a material with $n=2$ and on arranging the bands to have an appropriate density of states.

Three orbital models arise in a number of other  physically important contexts, including doped C$_{60}$ where, however, recent results suggest that lattice structure effects may be important.\cite{Takabayashi09} Extending our results to models with more realistic densities of states is a high priority for future research.

{\it Acknowledgements} AJM and CK were supported by DOE ER-46169 and PW by SNF PP002-118866. The calculations were performed with a code based on ALPS.\cite{ALPS}

\end{document}